# Brain Electrical Stimulation for Animal Navigation

**Amirmasoud Ahmadi[1], Sepideh Farakhor Seghinsara[1], Mohammad Reza Daliri[2] ,and Vahid Shalchian[3*]**

[1] M.Sc. Student, Biomedical Engineering Department, School of Electrical Engineering, Iran University of Science and Technology, Tehran, Iran

[2] Associate Professor, Biomedical Engineering Department, School of Electrical Engineering, Iran University of Science and Technology, Tehran, Iran

[3] Assistant Professor, Biomedical Engineering Department, School of Electrical Engineering, Iran University of Science and Technology, Tehran, Iran

___________________________________________________________________________________

## Abstract

The brain stimulation and its widespread use is one of the most important subjects in studies of neurophysiology. In brain electrical stimulation methods, following the surgery and electrode implantation, electrodes send electrical impulses to the specific targets in the brain. The use of this stimulation method is provided therapeutic benefits for treatment chronic pain, essential tremor, Parkinson's disease, major depression, and neurological movement disorder syndrome (dystonia). One area in which advancements have been recently made is in controlling the movement and navigation of animals in a specific pathway. It is important to identify brain targets in order to stimulate appropriate brain regions for all the applications listed above. An animal navigation system based on brain electrical stimulation is used to develop new behavioral models for the aim of creating a platform for interacting with the animal nervous system in the spatial learning task. In the context of animal navigation the electrical stimulation has been used either as creating virtual sensation for movement guidance or virtual reward for movement motivation. In this paper, different approaches and techniques of brain electrical stimulation for this application has been reviewed.

**Keywords:** *Rat Robot, Brain Computer Interface, Electrical Stimulation, Cyborg Intelligence, Brain to Brain Interface*

___________________________________________________________________________________

[*]**Corresponding authors**
**Address:** Neuroscience and Neuroengineering Research Laboratory, Biomedical Engineering Department, School of Elechtrical Engineering, Iran University of Science and Technology (IUST), Tehran, Iran
**Tel:** *+98-21-73225738, +98-21-73225628*
**Fax:** *+98-21-73225777*
**E-mail:** daliri@iust.ac.ir*, shalchyan@iust.ac.ir*

# تحریک الکتریکی مغز برای هدایت و جهت‌دهی حیوان

امیرمسعود احمدی[1]، سپیده فراخور سقین سرا[1]، محمدرضا دلیری[2]*، وحید شالچیان[3]*

[1]دانشجوی کارشناسی‌ارشد مهندسی پزشکی، گروه بیوالکتریک، دانشکده مهندسی برق، دانشگاه علم و صنعت ایران، تهران

[2]دانشیار، گروه بیوالکتریک، دانشکده مهندسی برق، دانشگاه علم و صنعت ایران، تهران

[3]استادیار، گروه بیوالکتریک، دانشکده مهندسی برق، دانشگاه علم و صنعت ایران، تهران

___

## چکیده

یکی از موضوعات مهم در بحث فیزیولوژی عصبی، موضوع تحریک الکتریکی مغز و کاربردهای گسترده آن می‌باشد. در روش‌های تحریک الکتریکی مغزی، پس از اجرای جراحی و کاشتن الکترود، پالس‌های الکتریکی به سمت نقاط مشخص مغز ارسال می شود. این روش تحریک، مزایای درمانی برای کنترل دردهای مزمن، کنترل رعشه، کنترل بیماری پارکینسون، کنترل افسردگی و همچنین کنترل اختلال حرکتی‌عصبی دارد. یکی از زمینه هایی که در آن اخیراً پیشرفت های خوبی صورت پذیرفته است، کنترل حرکت و جهت‌دهی حیوانات در مسیر خاص می باشد. در همه موارد نام برده شده شناسایی نقاط هدف به منظور تحریک مناسب در ناحیه از مغز از اهمیت زیادی برخوردار است. سیستم جهت‌دهی حیوان بر مبنای تحریک الکتریکی، به منظور توسعه مدل‌های رفتاری جدید با هدف ایجاد یک بستر ارتباطی با سیستم عصبی حیوان در فرآیند یادگیری حرکت در موقعیت های مکانی استفاده می‌شود. از تحریک الکتریکی مغز هم بعنوان راهنمایی با ایجاد حس مجازی و هم بعنوان عامل انگیزشی با ایجاد حس پاداش مجازی برای هدایت و جهت دهی حیوان استفاده شده است. در این مقاله، مروری بر انواع رویکردها، اصول و روش های تحریک‌الکتریکی مغزی استفاده شده در این کاربرد انجام گرفته است.

**کلیدواژه‌ها:** رتربات، واسط مغز و کامپیوتر، تحریک الکتریکی، هوش سایبرگ، واسط مغز-مغز

*محمدرضا دلیری- وحید شالچیان

**نشانی:** آزمایشگاه علوم و مهندسی اعصاب، گروه مهندسی پزشکی، دانشکده مهندسی برق، دانشگاه علم و صنعت ایران، تهران، ایران.

**تلفن:** ۷۳۲۲۵۶۲۸- ۷۳۲۲۵۷۳۸

**دورنگار:** ۷۳۲۲۵۷۷۷

**پست الکترونیکی:** daliri@iust.ac.ir، shalchyan@iust.ac.ir

## ۱- مقدمه

مفاهیم اولیه سیستم واسط مغز و کامپیوتر[1] اولین بار به عنوان یک مسیر مصنوعی از سیستم عصبی در سال ۱۹۶۴ توسط دکتر والتر معرفی شده است [۱]. امروزه، ترکیبی از سیستم های سخت افزاری و نرم افزاری برای برقراری ارتباط با سیستم عصبی را واسط مغز و کامپیوتر می نامند و تا به امروز، بیشتر تکنیک های واسط های مغز و کامپیوتر بر روی رمزگشایی فعالیت مغز تمرکز کرده اند [۱-۴،۵،۷]. به دلیل آنکه در اکثر سیستم های واسط مغز و کامپیوتر هدف اولیه ارائه روش های جایگزینی برای کمک به افراد معلول می باشد [۳]. به منظور درک سیگنال های مغز، انواع روش های تشخیصی و ثبت از جمله الکتروانسفالوگرام، الکتروکورتیکوگرافی[3]، مگنتوانسفالوگرام[4] و ثبت پتانسیل عمل و پتانسیل میدانی محلی از نورون ها[5] گسترش یافته است [۵-۷]. بخشی از سیستم های واسط مغز و کامپیوتر شامل تلاش هایی برای ایجاد ارتباطی میان مغز حیوانات و کامپیوتر می باشد. از جمله این مطالعات می توان به [۸، ۹] اشاره کرد که با رمزگشایی از سیگنال های پتانسیل میدانی محلی، میزان نیروی وارده از طریق دست رت را تخمین می‌زنند. در دو دهه اخیر، بیو ربات ها در انواع مختلف موجوداتی مانند حشرات [۱۰، ۱۱]، کبوتر [۱۲، ۱۳]، مارمولک [۱۴]، رت[۱۵-۱۷] ارائه شده است . با وجود اینکه تلاشهای گسترده ای در حوزه واسطهای مغز-کامپیوتر برای ترجمه سیگنالهای مغزی به دستورات کنترلی انجام شده است، رویکرد دیگری از تحقیقات ایجاد یک واسط برای ارتباط با مغز و تاثیر گذاری از طریق تحریک الکتریکی بافتهای مغزی برای مقاصد مختلف می باشد. در میان این تلاش ها، مطالعه [۱۶] ، توجه بسیاری از محققان را جلب کرده است. آنها نشان داده اند که حیوان را می توان از راه دور با میکروتحریک نواحی خاصی از مغز، هدایت و کنترل نمود. ناحیه دسته پیش مغز میانی[7] (MFB) به عنوان ناحیه هدف پاداش در نظر گرفته شده است که با مسیر دوپامینی مزولیمبک[8] مرتبط می باشد. از آنجائی‌که برای آموزش یک رفتار مطلوب در یک حیوان شرطی سازی بایستی انجام شود، برای یک آموزش موفق نیاز به انجام یک رفتار خاص و یک پاداش به‌دنبال آن رفتار است. بعنوان یک نمونه از آزمایشات رفتاری، لی و همکاران اثر دو ماز مختلف، یکی T شکل و دیگری مارپیچ W، را جهت آموزش رت‌ها استفاده کردند و بعد از پانزده مرحله آموزش نشان دادند رت هایی که توسط ماز T آموزش یافتند جواب بهتری به تحریک دادند و ماز ساده T شکل برای آموزش به مراتب بهتر از مارپیچ W بود [۱۷]. سان و همکاران روش اتوماتیک کنترل و هدایت حرکات رت را بررسی نمودند. در این پژوهش دستور تحریک، توسط واسط مغز و کامپیوتر به طور مستقیم به مغز رت اعمال شد و از شبکه عصبی رگرسیون عمومی[9] برای تجزیه و تحلیل و مدلسازی از روش کنترل انسانی استفاده شد و داده ها از موقعیت رت ، توسط دوربین های بالای ماز ، به کامپیوتر منتقل و توسط شبکه عصبی رگرسیون عمومی آنالیز شده و دستور اعمال می‌گردید [۱۵]. کنترل دقیق زمانی و پارامتر های تحریک در حلقه بسته می تواند بهبود قابل توجهی در روش کنترل اتوماتیک داشته باشد. در [۱۶] با استفاده از سخت افزار مناسب، توانسته اند تحریک و ثبت را از همان الکترود کاشته شده انجام دهند و به طور قابل ملاحظه ای تحریک های مصنوعی را کاهش دادند و به طور خاص این سیستم فیدبک تحریک در پاسخ به پتانسیل عمل را در پانزده ثانیه اعمال می کرد و همچنین تحریک را براساس وقایع رفتاری مانند ردیابی ویسکر رت که توسط ویدیو سرعت بالا ،گرفته می شد اعمال می کرده است. سیستم دیگر پیشنهاد شده است که تفاوت عمده ایی با سیستم های قبلی دارد: اولا این اجازه را به رت می دهد توسط دوربینی که در پشتش قرار دارد به نکاتی که انسان مشخص می کند مانند چهره انسان یا نشانه ها را دریابد و آنرا به کامپیوتر منتقل کند. دوماً می تواند رت را توسط نشانه ها به طور اتوماتیک هدایت کند (ابتدا پردازش تصویر بعد از اعمال تحریک) . اما مشکل هدایت اتوماتیک این است که یک تحریک برای چرخش به تنهایی نمی تواند باعث چرخش رت شود و باید یک سری تحریک اعمال شود که این مشکل با تقلید از هدایت مدل انسانی با حلقه بسته طراحی شد و این سری رت‌ها با کنترل اتوماتیک را به عنوان "سایبورگ رت[10]" نام گذاری کرده‌اند و نتایج نشان می‌دهد که این روش کنترل هدایت، درصد دقت و سرعتی همانند هدایت‌دستی دارد و حتی در بعضی موارد بهتر هم بوده است [۱۸]. اخیرا در [۱۹] به مقایسه ی روش های مختلف در حل ماز

---

پرداخته شده است: سه روش حل ماز : ۱- توسط کامپیوتری با تست مسیر های پیش رو ۲- توسط سایبورگ رت ۳- رت های معمولی. به رت های معمولی که به خط پایان می رسیدند پاداش کره بادام زمینی و آب داده می شد. در سایبورگ رت ها که به خط پایان می رسیدند ناحیه MFB تحریک می شده است. نشان دادند تعداد دفعاتی که سایبورگ رت از خانه های تکراری عبور کرده کمتر از رت های معمولی بود و تقریبا معادل کامپیوتر بوده است. پس حل ماز با معیار تعداد قدم‌ها ، در سایبورگ رت بهتر از رت های معمولی و در حد کارائی کامپیوتر بود. حد پوشش، یعنی تعداد خانه هایی که حداقل یک بار عبور داشته در سایبورگ رت ها بهتر از رت های معمولی و در حد کارائی کامپیوتر بود و عبور کمتری داشته است. زمان رسیدن به هدف تعیین شده در سایبورگ رت ها خیلی کمتر از رت های معمولی بود و در مجموعه در حل نمودن ماز، سایبورگ رت ها بهتر از رت‌های معمولی بودند. در پژوهش دیگری برای حل ماز فقط بر مبنای تحریک MFB استفاده شده در این کار آموزش سنگین تری داشته است و از رت های معمولی برای درک چرخش اشتباه استفاده شده است. در این کار رت های معمولی در یک ماز T شکل آموزش داده شدند. به نحو خاصی تحریک MFB اعمال می‌شود و زمانی‌که رت های معمولی مسیر اشتباهی را انتخاب می‌کردند تحریک قطع شده و رت های معمولی از این قطع شدن تحریک به اشتباه خود پی می‌بردند و در تست نهایی رت های معمولی توانسته اند با دقت میانگین نود درصد ماز را حل کند [۲۰]. می‌توان با کمک یک شبکه حسگر بی‌سیم، یک حیوان را برای عملیات جستجو و شناسایی محیط و همچنین در عملیات نجات استفاده کرد. هر یک از حیوانات می‌توانند یک شبکه حسگر بی‌سیم برای جمع‌آوری اطلاعات، پردازش، و ارتباطات شبکه و همکاری در مسیریابی و انتقال و بسته بندی، در کوله پشتی حمل نمایند. محققین یک شبکه حسگر را معماری نمودند و به عنوان یک طرح مسیریابی ساده اما کارآمد و مناسب به عنوان "کوله پشتی با قابلیت اکتشاف و جستجو" که می‌تواند تصاویر را دریافت و ارسال کند برای بهره برداری معرفی کردند [۲۱]. در این مقاله مروری بر رویکردهای مختلف، اصول و روش های تحریک الکتریکی مغز برای هدایت و جهت دهی حیوان خواهیم داشت و رویکردهای جدید در هدایت و جهت‌دهی رت مورد بررسی قرار گرفته و جمع‌بندی و مقایسه بین روش‌ها صورت پذیرفته است.

## ۲- نواحی اعمال تحریک

با تنظیم فعالیت های نورونی نواحی خاص می‌توان رفتار حیوانات را کنترل نمود و باتوجه به مطالعات سه رویکرد متفاوت در این زمینه وجود دارد:

۱- ایجاد انگیزه حرکت بر مبنای پاداش مجازی
۲- القای جهت حرکت بر مبنای حس مجازی [۱]
۳- تحریک مستقیم نواحی حرکتی

آماده‌سازی اولیه به همراه آموزش می‌تواند از راه‌های مختلفی از جمله پاداش و نشانه و یا احساس ترس انجام شود. رایج‌ترین روش برای این کار تحریک نواحی پاداش از جمله MFB، هیپوتالاموس و... می‌باشد. اگر تحریک پاداش همراه با رفتار صحیح حیوان باشد مراحل بعدی آزمایش و تست نهایی را حیوان سپری می‌کند و این فرآیند یک روش تنظیم فعالیت نورونی غیراجباری برای کنترل حرکت می‌باشد که در آن حیوان براساس اراده خود روند هدایت را طی می‌کند. از سوی دیگر، تنظیم فعالیت نورونی اجباری اغلب بوسیله تحریک نواحی مغزی مرتبط با حرکت صورت می‌پذیرد.

### ۱-۲- تحریک نواحی پاداش

در میان مناطق مختلف شناخته شده مغز، بسیاری از محققان برای اعمال شرایط پاداش مسیر دوپامینرژیک[۲] به عنوان ناحیهٔ هدف تحریک برگزیده اند. مسیرهای دوپامینرژیک، پیوندهای قوی شامل نورون‌های دوپامینرژیک است که دو ناحیه مغز را بهم وصل می‌کند. نورون دوپامینرژیک، دوپامین را به عنوان یک انتقال‌دهنده عصبی در مقصد سیناپسی خود انتقال می‌دهد. مسیرهای دوپامینرژیک نقش مهمی در ناحیه کنترل موتوری، رفتار مبتنی بر پاداش و انتشار هورمون دارند. بسته به ناحیه هدف، مسیرهای دوپامینرژیک در چهار مسیر متفاوت قرار می‌گیرند: مسیر مزولیمبیک[۳] مسیر مزوکورتیکال[۴]، مسیر توبِرو اینفوندیبولار[۵]. مسیر مزولیمبیک به عنوان یک مدار پاداش در میان فرآیندهای عصبی شناخته می‌شود [۲۲]. مسیر مزولیمبک از منطقه تگمنتال شکمی[۶] از طریق اگومبنس[۷] گسترش می‌یابد که شدیدا به یادگیری، حافظه، پاداش و انگیزه مرتبط می‌باشد. بنابراین به طور معمول به عنوان وسیله‌ای برای

---

[۱] Step
[۲] Dopaminergic
[۳] Mesolimbic
[۴] Mesocortical
[۵] Tuberoinfundibular
[۶] Ventral tegmental area (VTA)
[۷] Accumenbens

آموزش حیوانات مانند جوندگان و میمون استفاده می‌شود. ناحیه هدف معمول مسیر مزولیمبیک، ناحیه MFB است که شامل هسته اکومبنس می‌باشد. جمیع محققان پذیرفته‌اند که تحریک ناحیه MFB باعث احساس پاداش و لذت می‌شود و اثر پاداش تحریک آن با استفاده از دوز مناسبی از آنتاگونیست‌های دوپامین تنظیم می‌شود [۲۳]. از آنجایی که فعال شدن مسیر مزولیمبیک باعث احساس مطلوب و پاداش می‌شود، بسیاری از گروه‌های تحقیقاتی از تحریک این ناحیه برای آموزش رت استفاده می‌کنند. با این حال، برای کنترل و هدایت حیوانات تحریک این ناحیه به تنهایی کافی نمی‌باشد و باید همراه با علائم حرکتی اعمال شود. با این وجود، تالوار و همکاران گزارش داده‌اند که تحریک MFB نه‌تنها برای حرکت به سوی جلو بلکه برای حرکت‌های ابتکاری مانند صعود یا فرود نیز می‌تواند مورد استفاده قرار گیرد [۱۶]. اتاقک شرطی سازی معمولا به نام "جعبه اسکینر" شناخته می‌شود و توسط اسکینر برای مطالعات رفتاری حیوانات به ویژه برای شرطی سازی عامل و شرطی سازی کلاسیک طراحی شده است و به‌طور گسترده‌ای برای آموزش پاداش MFB در جوندگان استفاده می‌شود [۲۴]. ترس، یکی از احساساتی است که توسط مدولاسیون نورونی کنترل می‌شود که در حیوانات به صورت غریزی در طول تکامل ایجاد شده است. یادگیری برپایه ترس بسیار قوی‌تر از یادگیری مبتنی بر پاداش است. سیستم لنفاوی به احساسات مانند لذت، خشم، ترس و شادی در مغز مرتبط است. سیستم لیمبیک شامل هیپوکامپ، هیپوتالاموس، آمیگدال می‌باشد [۲۷]. آمیگدال شامل دو گروه هسته است که در انتهای سیستم لیمبیک قرار دارد. هوآی و همکاران برای کنترل حرکت رت ناحیه آمیگدال را هدف قرار داده‌اند و با تحریک‌الکتریکی این ناحیه توانسته‌اند رت را کنترل نمایند و رت را براساس غریزه فرار از مجازات که به‌صورت مجازی اعمال می‌شود، هدایت کرده‌اند [۲۸]. با این حال برخی از مطالعات نشان داده‌اند که آمیگدال نقش محوری در تمایز میان پاداش‌ها بازی می‌کند [۲۹]. کوئن و پرزکوت یک سیستم کنترل و هدایت رت را براساس مجموعه نواحی گانگلیاهای هیپوکامپ-آمیگدال-بارل طراحی کرده‌اند [۳۰]. انتخاب ناحیه آمیگدال برای مجازات مجازی مورد بحث است و به‌همین دلیل به‌عنوان یک جایگزین احتمالی، ژنک و همکارانش، ناحیه خاکستری پریاکتونیک را

پیشنهاد داده‌اند که ترکیبی از ترس و اضطراب را القا می‌کند و برای دستور "توقف" در عمل هدایت رت مورد استفاده قرار گرفته‌است [۳۱, ۳۲]. تحریک ناحیه خاکستری پریاکتونیک باعث رفتار تدافعی و همچنین باعث افزایش فشارخون و ضربان قلب می‌شود [۳۳-۳۵].

## ۲-۲- تحریک ناحیه حسی‌پیکری

تحقیقات در زمینه کنترل حیوانات، مخصوصاً جوندگان، با تحریک‌الکتریکی در مغز، توسط چندین محقق انجام شده‌است. برخلاف ناحیه‌های موتوری که باعث حرکت غیرارادی می‌شود، تحریک قشرحسی‌پیکری بیشتر مربوط به آموزش حرکت‌های داوطلبانه حیوان براساس خواسته‌ی آزمایش‌کننده می‌باشد. در فاز آموزش حرکت از تنظیم زمانی تحریک MFB و تحریک ناحیه حسی‌پیکری استفاده می‌کنند. تحریک ناحیه حسی‌پیکری در نیم‌کره راست باعث می‌شود که رت در سمت مقابل (سمت‌چپ) احساس برخورد به یک مانعی را داشته باشد. همین امر برای تحریک بر نیم‌کره چپ نیز صادق است. قشر حسی پیکری موجود در هردو نیم‌کره راست و چپ به عنوان نشانه به حیوان استفاده شده‌است و اعتقاد برآن است پاسخ به تحریک دوطرفه است به این منظور که تحریک یک نیم‌کره بر دیگری تأثیر می‌گذارد [۳۶]. به‌طورخاص، رت‌ها اطراف خود را به‌وسیله ویسکرهای خود شناسائی می‌کنند. دریافته‌اند که حتی فعالیت یک نورون این ناحیه می‌تواند برروی شناسائی محیط حیوانات تأثیر بگذارد [۳۷]. ناحیه مربوط به ویسکر یا قشر بارل یکی از مناطق معمول برای راهنمائی جهت حرکت به رت می‌باشد [۱۵, ۱۶, ۱۹, ۲۰, ۳۸, ۳۹]. فاصله بین ویسکر ها بسیار کم می‌باشد اما در مغز به شکل بهتری نواحی مرتبط به‌آنها توزیع شده‌است (شکل ۱). به همین علت برای کنترل جهت حرکت مورد استفاده قرار می‌گیرد [۴۰]. با آنالیز سیگنال‌های الکتریکی در قشر بارل در زمان هدایت رت در یک دالان دریافته‌اند که قشر بارل رت اطلاعات مفیدی در ارتباط با حرکت و فاصله در خود ذخیره می‌کند [۴۱]. رت محیط اطراف خود را به وسیله مجموعه ویسکرهای خود با پردازش محرک‌های محیطی می‌شناسد. بسیاری از مطالعات توصیف کرده‌اند که مسیر نیگرواستریاتال ارتباط عمیق با رفتار چرخشی در حیوانات دارد [۴۲-۴۴].

## ۲-۳- تحریک مسیر موتوری

---

[1] Dopamine antagonists
[2] Operant conditioning
[3] Neuromodulation
[4] Limbic System
[5] Periaqueductal Gray
[6] Somatosensory

طرح‌هایی که مسیر موتوری را به‌جای نواحی پاداش به عنوان هدف تحریک در نظر می‌گیرند، هدایت را خیلی مستقیم‌تر انجام می‌دهند. با توجه به سلسله مراتب سیستم موتوری،

باشد. با این‌حال بیشتر تحقیقات این حوزه بر توان بخشی عصبی تمرکز نموده‌اند [46-48].

**جدول(1) : مشخصات کلیدی مدالیته‌های مختلف**

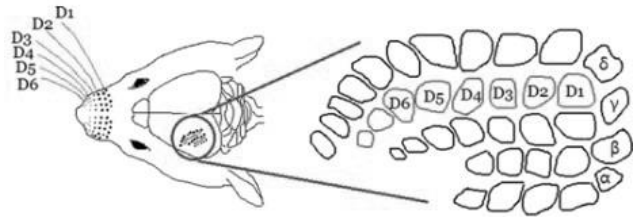

شکل(1) – قشر بارل در رت [50].

اهداف زیادی برای کنترل حرکت حیوانات وجود دارد و سیستم موتوری به طور کلی سه سطح کنترل دارد: اولاً سطح سیگمنتال[1]، دوماً سطح پیش‌بینی، سوماً سطح پیش حرکتی [45]. در مرحله اول، سطح سیگمنتال کمترین سطح در سلسله مراتب موتوری دارا می‌باشد و سازماندهی مدارهای نخاع که حرکات اتوماتیک گروه‌های خاصی از فیبرهای عضلانی از جمله حرکت و برخی از فعالیت‌های موتوری تکراری را کنترل می‌کند. برخی از رویکردهای حرکت برمبنای هدف قراردادن نخاع می-

## 3- اصول و روش‌های تحریک‌الکتریکی مغز

برای کنترل حرکت حیوانات چندین منطقه مغز مورد هدف قرار گرفته‌است. تحریک ولتاژ و جریان دو روش معمول تحریک الکتریکی می‌باشد. درمیان مهندسان عصبی، عملکرد قطار های پالس دوقطبی[2] مؤثرتر از قطارهای پالس تک‌قطبی برای تحریک عصبی گزارش شده‌است [49]. در میان شکل موج‌های مختلف، شکل موج مستطیلی معمولاً برای تحریک مغز به‌کار می‌رود و اثرات تحریک متناسب با اینکه تحریک از نوع ولتاژ یا جریان باشد، متفاوت است. پالس‌های تک‌قطبی تنها یک قطب مثبت یا منفی با فاز صفر دارد و پالس دوقطبی دارای فاز مثبت و منفی است و پارامترهای تحریک شامل چهار عنصر دامنه، عرض پالس، نرخ پالس و زمان کل تحریک می‌باشد. شدت تحریک را می‌توان با تنظیم دامنه و زمان پالس کنترل کرد. شکل2، شکل موج تک‌قطبی و دوقطبی را توصیف می‌کند و برای درک بهتر، روش انتقال محرک تولید شده نیز ارائه شده-است. پالس تک‌قطبی از آنجایی که تنها یک قطب دارد، منجر به واکنش برگشت ناپذیر فارادیک و افزایش احتمال آسیب رسیدن به بافت یا الکترود می‌شود. همانطور که در شکل 3 نشان داده شده است هنگامی که با قطار تک قطبی عمل تحریک صورت می‌پذیرد، بارهای منفی در محل تحریک باقی می‌ماند زیرا هیچ فازی برای لغو بار منفی اعمال نمی‌شود و این امر می‌تواند منجر به آسیب به بافت‌های اطراف و همچنین الکترود شود. در پالس دوقطبی، یک جریان ثابت در یک الکترود در یک جهت و به الکترود دیگر با فاز معکوس منتقل می‌شود. فاز اول که فاز اصلی است فاز تحریک نامیده می‌شود که باعث تولید پتانسیل عمل می‌شود و مرحله بعد فاز معکوس است که معکوس فعالیت الکتروشیمیایی در این فاز اتفاق می‌افتد که باعث کاهش فشار بار در بافت‌های اطراف می‌شود [50]. (شکل4) با این حال، همانطور که در شکل 4 دیده می‌شود، حتی اگر میزان بار منفی و مثبت متعادل باشد پتانسیل اعمال شده به ناحیه تحریک همزمان با کاهش بار منفی، افزایش می-یابد و ممکن است باعث خوردگی الکترود شود و به‌منظور

| نواحی تحریک | پارامترهای تحریک نواحی مختلف |
|---|---|
| پاداش | فاصله پالس‌ها در اکثر مقالات 10 میلی‌ثانیه می‌باشد که مدت زمان پالس 1 میلی‌ثانیه و تعداد پالس‌ها بین 10 تا 15 عدد و دامنه پالس‌ها 1 تا 10 ولت‌اند [18] [20]. در یکی از مقالات نیز ازپالس‌هایی با فاصله 4 میلی‌ثانیه که دامنه پالس 0.5 تا 0.2 میلی‌آمپر و مدت زمان پالس 0.2 میلی‌ثانیه متغیر بوده استفاده شده است [17]. |
| حسی-پیکری | تعداد پالس‌های ارسال شده در این ناحیه از 5 تا 15 عدد با توجه به پارامترهای دیگر قابل تغییر بوده است. در اکثر مقالات زمان پالس ارسالی 1 میلی‌ثانیه و دامنه 6 تا 8 ولت می‌باشد. قطار تحریک 0.2 هرتز با فرکانس 100 هرتز در مدت زمان پالس 0.5 میلی ثانیه نیز اعمال شده است [18] [20] [77]. |
| موتوری | برای ناحیه پستر‌امدیکال شکمی از پالس‌های مربعی که دارای فرکانس 50 هرتز و مدت زمان پالس 5 میلی‌ثانیه و مدت زمان قطار پالس 0.5 ثانیه استفاده شده است [78]. در مسیر نیگرواستریاتال پالس‌ها دارای مدت زمان 0.2 میلی ثانیه با فاصله پالس 4 میلی ثانیه و دامنه 200 میکروآمپر اند [39]. |

---

[2] Biphasic Pulse            [1] Sigmental

جلوگیری از این اتفاق دو روش پیشنهاد شده‌است: روش اول این است که مجموعه سطح دامنه‌های مختلف برای فاز مثبت و منفی برای جلوگیری افزایش پتانسیل در زمان قطع بار (شکل۵) و روش دوم این است که فاز صفر-دامنه را بین فازهای منفی و مثبت وارد کنیم، زمان کافی برای بازیابی از شارژ و تخلیه فراهم شود (شکل۶) [۵۰].

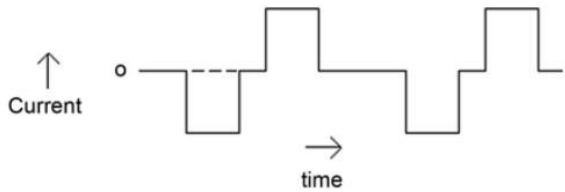

شکل(۶)- پالس دوقطبی با شارژ متوازن و تاخیر بین فاز [۵۰].

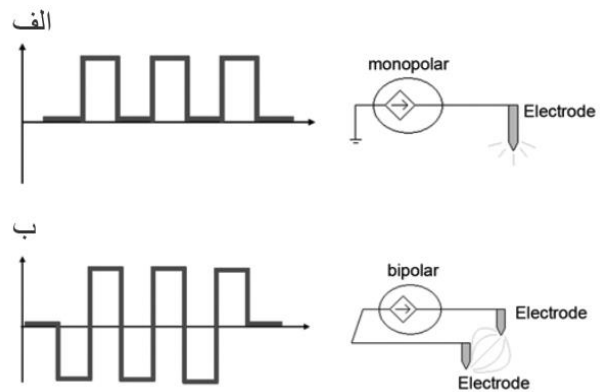

شکل(۲)- اصول روش تحریک. الف) تک‌قطبی. ب) دوقطبی[۵۰].

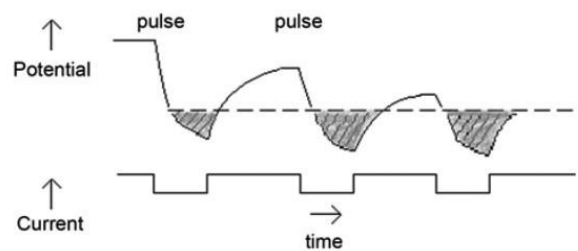

شکل(۳)- پالس تحریک تک‌قطبی[۵۰].

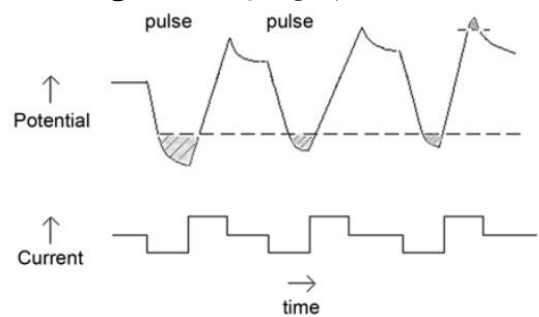

شکل(۴)- پالس تحریک دوقطبی [۵۰].

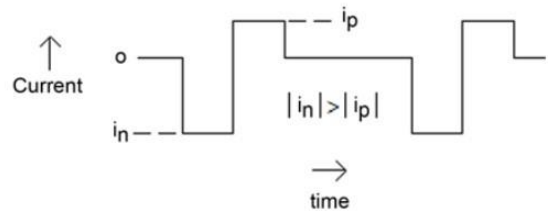

شکل(۵)- پالس دوقطبی با شارژ نامتوازن[۵۰].

بدن انسان دارای خصوصیت مقاومت-خازنی وابسته به فرکانس می‌باشد در روش تحریک تک‌قطبی تخلیه جریان ورودی مغز بعداز انتشار در محیط مغز توسط مرجع صورت می‌پذیرد و باعث آسیب به بافت‌های مغز می‌شود اما در تحریک دوقطبی موج مربعی، تخلیه بسیار سریع اتفاق می‌افتد و تحریک متمرکزتر بر نقطه هدف اعمال می‌شود زیرا جریان خروجی توسط الکترود قطب مخالف هدایت می‌شود همچنین تغییر فوری میدان الکتریکی از منفی به مثبت باعث افزایش گرادیان الکتریکی می‌شود که موجب دفع NA از طریق غشاء شده و gNa افزایش می‌یابد بنابراین حد آستانه برای دیپالیزاسیون تأمین شده و آسکون‌های بیشتری فعال می‌شوند در نتیجه تحریک دوقطبی نسبت به تحریک تک‌قطبی به منظور دستیابی به آستانه حسگرایانه نیاز به ولتاژ پیک کمتری دارد [۵۲, ۵۱]. تحریک کاتدی منجر به دپلاریزاسیون و تحریک آندی سبب هایپرپلاریزاسیون محلی غشا می‌گردد. با توجه به قوانین کولن مناطق هایپرپلاریزاسیون بطور غیرمستقیم در بخش‌های دورتر از غشا دپلاریزه می‌گردند بنابراین تحریک آندی در هایپرپلاریزاسیون محلی و بطور غیرمستقیم در دپلاریزاسیون اجسام سلولی دورتر نتیجه می‌دهد. ازاین‌رو تحریک کاتدی محلی و تحریک آندی دورتر از سلول هدف باعث ایجاد دپلاریزاسیون و فعالیت سلول‌ها در لایه‌های مغز می‌گردد [۷۹]. همچنین به‌این صورت توجیه نموده‌اند که آکسون در مناطقی که گرادیان ولتاژ ( $E = -dv/dr$ ) حداکثر است، دیپولاریزه می‌شود. تراکم بالا پتانسیل اکثراً در گوشه‌ها و لبه‌ها در مجاورت الکترود اتفاق می‌افتد و به‌صورت کلی کمترین حد دپولاریزاسیون زمانی اتفاق می‌افتد که آکسون عمود بر خطوط پتانسیل عبور می‌کند. معادله حاکم بر $E_z$ و $E_R$ برای الکترود تک‌قطبی و دوقطبی در زیر آمده است (فرمول(۲,۱) ). در این فرمول $\rho$ مقاومت در برابر بافت و فاصله بین دو الکترود D می‌باشد. گسترش جریان (r=f(I)) را می‌توان برای هر دو حالت پیش‌بینی نمود که I نمایش‌دهنده مقدار کل جریان عبوری از الکترود است و براساس معادلات مغناطیسی و شواهد تجربی

نتیجه گرفته‌اند که اولاً، در جریان بسیار کم (0.1mA~) تحریک دوقطبی گسترش بهتری می‌یابد ولی در جریان‌های بالا گسترش جریان کم می‌باشد و این واقعیت به این دلیل است که با توان سوم فاصله مقدار جریان کاهش می‌یابد. دوماً برای جریان‌های بالا (10mA~) استفاده از تحریک تک‌قطبی مفیدتر می‌باشد [۸۰].

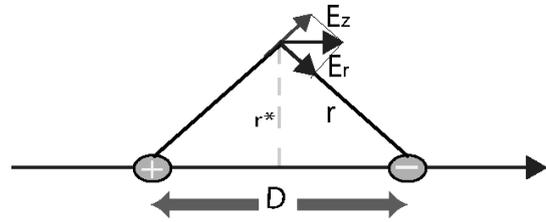

**شکل(۷)- مدل مورد استفاده برای محاسبه گرادیان ولتاژ**

$$E(r) = \frac{-dv}{dr} = \frac{\rho}{4\pi} * \frac{I}{r^2} \qquad (1)$$

$$E_z(r^*) = \frac{-dv}{dr^*} = \frac{\rho D}{4\pi} * \frac{I}{r^{*3}} \qquad (2)$$

همچنین تحقیقی که برروی انسان بر روی تفاوت این دو نوع از الکترودهای تحریک در بیماران پارکینسونی از نظر عوارض جانبی آن بررسی شده است و به این نتیجه رسیده‌اند که تحریک دوقطبی دارای تحریک متمرکزتری می‌باشد و ناحیه نازکتری از بافت فعال می‌شود [۸۱]. فیلدفوت و همکاران دریافته اند که برخلاف باورهای سنتی، فاز مثبت پس از فاز منفی اثر تحریک‌کننده بیشتری در تحریک شدت پایین دارد [۴۹]. به غیر از تحریک الکتریکی قشر حسی پیکری، روش‌هایی برای ایجاد نشانه‌دهی هدایت در سایر اشکال تحریک نظیر تحریک‌نوری وجود دارد. هابر و همکاران با ریز تحریک‌نوری قشر بارل توانسته‌اند رفتاری را به موش در حرکت آزادانه آموزش دهند و موش آموزش‌دیده تحریک نوری را تشخیص دهد [۵۳]. در زمان‌هایی که تحریک‌های تهاجمی ضرورتی ندارد، روش‌های غیر تهاجمی برای نشانه به حیوانات با استفاده از تحریک اپتوژنتیک اعمال می‌شود¹. سلچیدهاناندام و همکاران ناحیه بارل در رت را بوسیله روش اپتوژنتیک تحریک نموده‌اند و دریافته‌اند که آموزش اپتوژنتیک مبتنی بر پاداش می‌تواند در کنترل رفتار رت مورد استفاده قرار بگیرد [۵۴].

## ۴- الگوهای آموزش و هدایت

رت‌ربات یکی از معمولی‌ترین بیوربات‌ها می‌باشد [۱۶]. امروزه محققان درابتدا از سه دستورالعمل هدایت( راست، چپ، روبه‌جلو) برای هدایت یک رت‌ربات استفاده می‌کنند. آموزش فرآیند هدایت یکی از مراحل ضروری برای آموزش الگوی عملکردی به رت می‌باشد. در طی آموزش دستی، آموزش‌دهنده باید رت‌ربات را به صورت پیوسته مشاهده کند و دستورات کنترلی را از طریق تحریک‌الکتریکی مکرراً ارسال کند. آموزش دستی سه مشکل اصلی دارد. اولاً، آموزش دهنده باید در آموزش رت‌ربات حرفه‌ای باشد و برای افراد بی‌تجربه سخت می باشد که رت‌ربات را آموزش دهند. دوماً، مراحل یادگیری و رفتار رت‌ربات را نمی‌توان برای تجزیه‌وتحلیل کمی و تنظیم پارامتر ثبت کرد. برای مقابله با این مشکلات سیستم آموزش اتوماتیک را پیشنهاد داده‌اند. در این سیستم، رفتار رت‌ربات توسط یک دوربین تحت نظارت قرار گرفته و با استفاده از یک رایانه بصورت زمان‌واقعی² آنالیز می‌شود و براساس نتایج تحلیل شده، کامپیوتر به طور مدام تصمیم می‌گیرد تا فعالیت‌های آموزشی را رت‌ربات انجام دهد و یک چارچوب آموزش سلسله مراتبی که دارای یک لایه واکنشی و یک لایه مشورتی است معرفی می‌شود و براساس این چارچوب سیستم آموزش در زمان واقعی تحریک‌الکتریکی برای آموزش رت‌ربات را انجام می‌دهد [۱۸]. روش دیگری نیز با الگوبرداری و مدل‌سازی از هدایت انسانی عمل هدایت اتوماتیک را انجام می‌دهد [۱۵].

### ۱-۴- آموزش و هدایت دستی

تصویری از رت‌ربات در شکل۹ آمده است. یک جفت الکترود ریز تحریک در بسته نرم افزاری جلو مغز میانی و جفت دیگر در ناحیه چپ و راست حسی‌پیکری در مغز رت کاشته می‌شود. پس از پنج روز بهبودی، یک دستگاه تحریک بی‌سیم در پشت رت برای رساندن میکروتحریک تجهیز می‌شود و کاربر با استفاده از یک رایانه امکان ارسال پالس‌های محرک را برای هر یک از نواحی مغز حیوان از فاصله صد متری از طریق بلوتوث فراهم می‌کند. شدت پالس تحریک توسط تعداد پالس و دامنه پالس تعیین می‌شود. تحریک ناحیه بسته جلومغز میانی سطح دوپامین را بالا می‌برد و تحریک نواحی حسی پیکری باعث می‌شود که رت تصور کند به مانعی درسمت مخالف تحریک برخورد کرده است [۵۵, ۵۶]. اتاقک شرطی‌سازی اغلب برای آموزش یک رفتار مطلوب در یک حیوان استفاده می‌شود [۱۷].

---

¹ Optogenetic

² Real Time

اجزای اصلی اتاقک عبارتند از یک اهرم، بلندگو، چراغ سیگنال، غذا و یا آب و کف آن شبکه ایی فلزی برای شوک الکتریکی

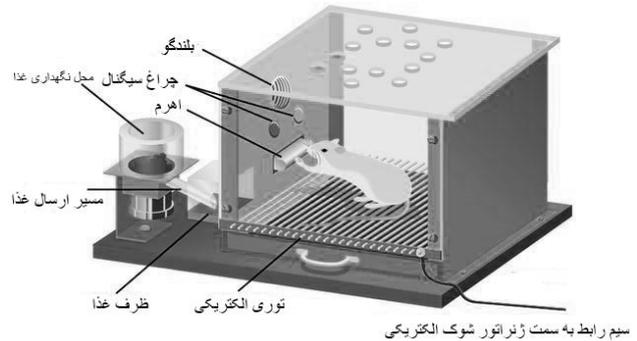

شکل(۸) - اتاقک شرطی ساز (جعبه اسکینر)

می‌باشد. آموزش با اتاق شرطی‌سازی عمیقاً مرتبط به مفهوم تقویت شرطی و مجازات است. چهار گروه اصلی وجود دارد: اول تقویت مثبت، دوم تقویت منفی، سوم تنبیه مثبت، چهارم تنبیه منفی است. باتوجه به تحریک دستجات پیش مغز میانی تقویت مثبت مطلوبترین توصیف می‌باشد. تقویت مثبت زمانی اتفاق می‌افتد که پاسخ رفتاری رت در راستای رویداد مطلوب یا محرک اعمال شده‌باشد [۲۵]. تقویت مثبت به تحریکی گفته می‌شود که برای رفتار خاصی به حیوان اعمال می‌شود و در راستای آن هدف می‌باشد [۲۶]. یکی از روش‌های پاداش حاصل از فشردن اهرم در اتاقک شرطی‌سازی (شکل ۸) و اعمال تحریک الکتریکی به مغز می باشد که مطالعات صورت گرفته بر روی نواحی دستجات پیش مغز میانی و هسته عروقی خلفی و یا آمیگدال می باشد که ژنراتور تحریک در خارج اتاقک است و زمان‌بندی و پارامترهای تحریک را کنترل می‌کند. اگر الکترود در ناحیه دستجات پیش مغز میانی کاشته شود هربار که رت اهرم را فشار می‌دهد تحریک الکتریکی پاداش اعمال می‌شود. در طول این آزمایش، افزایش تمایل رت به معنای آموزش درست می‌باشد این افزایش میل به فِشردن اهرم به منزله احساس پاداش در حد مطلوب و مورد قبول می‌باشد. در امر آموزش رتربات، تحریک MFB به عنوان پاداش و همچنین هدایت "حرکت رو به جلو" استفاده می‌شود [۵۷, ۵۸]. تحریک چپ و راست حسی‌پیکری به عنوان نشانه برای چرخش به چپ و راست استفاده می‌شود. برای بدست آوردن پاداش، رتربات باید رفتارهای صحیح مرتبط با نشانه را آموزش ببیند. رتربات می تواند تحریک مغز را از راه دور به عنوان دستورالعمل جهت حرکت دنبال کند.

درطول آموزش، شدت تحریک‌های چرخشی بدون تغییر باقی می‌ماند و فقط به‌عنوان یک نشانه عمل می‌کند اما شدت (تعداد پالس و دامنه پالس) تحریک پاداش (FORWARD) با توجه به شرایط یادگیری برای فعال کردن رتربات افزایش می‌یابد و برعکس.

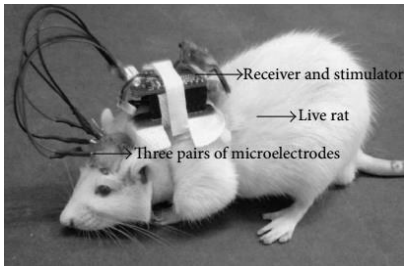

شکل(۹) - رتربات[۱۸].

### ۴-۱-۱- فرآیند یادگیری

روش آموزش دستی در شکل ۱۰ نشان داده شده‌است. الکترودهای ریزتحریک در عمل جراحی وارد مغز می‌شوند و فلش‌های فیدبک نشان‌دهنده نیاز به آموزش دوباره رتربات است. دو فرآیند که در مستطیل با نقاط نقطه‌چین قرار دارند مهمترین بخش آموزش هستند و در ادامه شرح جزئیات آنها آمده‌است. روش تنظیم پارامترهای تحریک بهینه برای رتربات شامل دو بخش است، بخش اول فشار دادن اهرم و بخش دوم تنظیم چپ/ راست است. اگر تحریک بیش از حد خفیف باشد رتربات را تحریک نخواهد کرد و اگر بیش از حد قوی باشد رتربات مجروح خواهد شد. مقدار مطلوب تحریک پاداش در فرآیند فشردن اهرم بهینه‌سازی می‌شود و همچنین مقدار مطلوب تحریک چرخشی در روش تنظیم چپ و راست پدیدار می‌شود. فرآیند تقویت شامل حرکت آموزش رتربات برای پاسخ صحیح به تحریک می‌باشد. محیط آموزشی مورد استفاده در این قسمت یک ماز هشت بازوئی است. در این فرآیند سه وظیفه داریم.

الف) حرکت رو به جلو ($T_0$) : این وظیفه معمولا به‌عنوان $T_0$ نام‌گذاری می‌شود. رتربات باید ماز هشت بازوئی را در جهت عقربه‌های ساعت یا خلاف عقربه‌های ساعت بدون هیچ چرخشی اشتباهی طی کند. ب) چرخش به چپ و چرخش به راست ($T_1$) : این وظیفه به عنوان $T_1$ نام‌گذاری می‌شود به این صورت است که تحریک چپ و پاداش، زمانی که اعمال می‌شود رتربات باید ماز هشت‌بازوئی را در جهت عقربه‌های ساعت بدون هیچ اشتباهی طی کند و تحریک راست و پاداش برعکس می‌باشد. براساس نحوه عمل جراحی، قبل از هر آموزشی بوسیله تحریک

---

[1] Ventral posterdateral nucleus

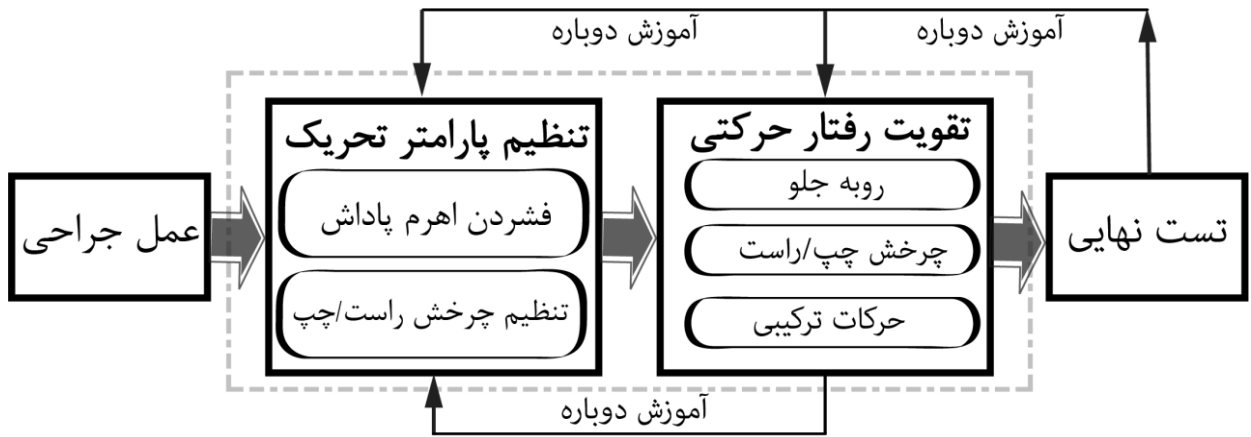

**شکل(۱۰) - الگوریتم آموزش هدایت به‌صورت دستی**

بچرخند در حالی‌که بعضی‌های دیگر تمایل به چرخش به سمت راست دارند. اگر برای مثال رتربات تمایل به چرخش به سمت چپ دارد در فاز $T_1$ عمل چرخش به سمت چپ در مرحله اول آموزش داده می‌شود و برعکس. $T_{1a}$ عمل آموزش چرخش به سمت چپ و $T_{1b}$ آموزش چرخش به سمت راست در این حالت می‌باشد.

ج) چرخش ترکیبی ($T_2$) : این وظیفه برای تقویت رفتارهای چرخش چپ و چرخش به سمت راست به‌صورت همزمان می‌باشد و به‌وسیله پاداش و نشانه های چرخشی رندوم عمل هدایت صورت می‌پذیرد. درطول تقویت رفتارهای حرکتی، آموزش دهنده ممکن است سطح تحریک پاداش را براساس تجربیاتش در آموزش هدایت رتربات تنظیم کنند و بعداز اتمام وظیفه $T_2$، رتربات برای هدایت آماده می‌باشد.

## ۴-۲- هدایت و آموزش به‌روش اتوماتیک

دو روش متفاوت در هدایت و آموزش اتوماتیک در مقالات پیشنهاد شده‌است. روش اول سعی بر آن دارد با شکل‌دهی قواعد ریاضی هدایت را ساماندهی نماید و روش دوم با استخراج اطلاعات و مدل‌سازی از هدایت دستی هدایت را انجام می دهد. رتربات ها دارای خودآگاهی می‌باشند و ممکن است تحت فشار فرآیند آموزش واکنش نامطلوب نشان دهند و همچنین بدن رتربات غیرقابل تشخیص است و موقعیت‌های مختلفی دارد. بنابراین، اختصاص آموزش خودکار به یک کامپیوتر با چهار مساله اساسی روبرو می‌باشد:

الف) چگونگی تنظیم وظایف در فاز آموزش برای اطمینان از اینکه رتربات برای هدایت واجد شرایط است.

ب) چگونگی شناسایی حرکت و رفتار رتربات در زمان واقعی.

ج) نحوه تشخیص حالت غیرطبیعی رتربات در فاز یادگیری مانند عدم واکنش یا واکنش بیش از حد.

د) توسعه یک استراتژی تحریک هوشمند که قادر به تحریک-الکتریکی انعطاف‌پذیر باشد.

### ۴-۲-۱- یادگیری وظائف

در مطالعه [۳۸] دریافته اند که آموزش حرکت در فرآیند فشردن اهرم و حرکت به‌سوی جلو ($T_0$) نیاز نمی‌باشد زیرا این رفتار در $T_1$ آموزش داده می‌شود. آموزش اتوماتیک شامل تنها $T_1$ و $T_2$ است. در کارهای گذشته دیده اند که بدون آموزش مرحله $T_1$ قبل از مرحله $T_2$ تعدادی از رتربات ها در آزمون نهایی موفق نبوده‌اند و این نشان می‌دهد که روند آموزش رتربات باید به ترتیب انجام شود[۲۰]. کل روند آموزش اتوماتیک در شکل ۱۱ نشان داده شده است. فرآیند تنظیم پارامترهای تحریک در بخش آموزش دستی توضیح داده شده‌است و فرآیند تقویت رفتار حرکتی که بیشترین زمان در فاز آموزش دستی را به خود اختصاص می‌دهد در این روش به‌صورت خودکار انجام می‌پذیرد [۱۸].

### ۴-۲-۲- چارچوب کلی

چارچوب آموزش خودکار سلسله مراتبی است که در شکل ۱۲ نشان داده شده است. ورودی تصاویری است که توسط یک دوربین از بالا از رتربات گرفته می‌شود و خروجی آن تحریک می‌باشد. این چارچوب دارای دولایه است، لایه اول، لایه واکنشی است که در زمان‌واقعی براساس مدل وظیفه و وضعیت دریافتی، راهنمائی های آموزشی را فراهم می‌کند. این لایه تصمیم می‌گیرد کدام تحریک ( چپ، راست، حرکت رو به جلو) در چه زمانی ارسال می‌شود. لایه دوم، لایه مشورتی[1] مسئول

---
[1] Deliberative

شکل(۱۱)- الگوریتم آموزش هدایت به‌صورت اتوماتیک

تنظیم تحریک پاداش متناسب با شرایط یادگیری رت‌ربات است. هر واحد این چارچوب در ادامه شرح داده شده است.

### ۴-۲-۲-۱- ردیابی رت

دو روش اصلی برای ردیابی موش وجود دارد یکی علامت‌گذاری سر و بدن رت به‌وسیله یک رنگ روشن و براق می‌باشد [۱۵] و روش دیگر قراردهی سنسور فشار زیر پای رت می‌باشد در این روش بدست آوردن موقعیت دقیق سر و بدن رت دشوار می‌باشد. روش‌های جدید و مرسوم در ردیابی رت‌ربات مبتنی بر جریان نوری برای بدست‌آوردن پارامترهای حرکت (موقعیت سر، موقعیت بدن و جهت سر) و حرکت رت‌ربات است.

**ردیابی حرکت:** دراین روش اولین مهم آن است که رت‌ربات در چه موقعیتی (موقعیت بدن و سر) می‌باشد و به کدام مسیر در محیط آموزش باید هدایت شود. جزئیات الگوریتم ردیابی حرکت به شرح زیر است:

**موقعیت بدن:** پس از حذف پس زمینه، یک مستطیل کوچک برای پوشش رت‌ربات در تصویر ورودی در نظر می‌گیرند و مستطیل دیگری که بیشترین پیکسل هدف را دارد ذخیره می‌شود. میانگین موقعیت‌ها، موقعیت پیکسل هدف به عنوان موقعیت بدن محاسبه می‌شود [۱۸].

**موقعیت سر:** هنگام ردیابی یک هدف متحرک، نقاط ویژگی مانند گوشه‌ها معمولا مورد استفاده قرار می‌گیرد [۵۹]. از آنجایی که رت‌ربات دارای یک کوله‌پشتی است بیشتر گوشه‌ها در اطراف منطقه کوله‌پشتی بسیار نزدیک به‌سر ظاهر می‌شود و میانگین موقعیت این گوشه‌ها را به‌عنوان موقعیت سر درنظر می‌گیرند. ویژگی‌ها را توسط الگوریتم تشخیص ویژگی Shi-Tomasi استخراج کرده‌اند [۶۰].

**جهت سر:** مسیر مابین موقعیت بدن و موقعیت سر به‌عنوان جهت سر رت‌ربات درنظر گرفته می‌شود. ادعا کرده‌اند که این روش با دقت نزدیک به ۹۰٪ ردیابی را انجام می‌دهد.

**ردیابی رفتار:** بعد از آنکه پارامترهای اولیه حرکت را دریافت کرده‌اند برای شهود رفتار رت‌ربات در محیط آموزش مانند عدم حرکت، مراقبت از بدن و بالا رفتن از دیوار پلکسی گلاس به‌کار می‌رود. در روش آموزش دستی زمانی که رت‌ربات بی‌حرکت می‌ماند فقط تحریک پاداش برای رت‌ربات ارسال می‌شود و هنگامی که یک رت‌ربات در حال مراقبت از خود و یا بالا رفتن[1] از دیوار است نباید هیچ تحریکی ارسال شود. با این حال، در سیستم آموزش اتوماتیک امکان شناسایی حالت‌های مراقبت و بالا رفتن به‌صورت اتوماتیک نمی‌باشد ( این دو حالت را به‌صورت دستی برای تجزیه و تحلیل رفتاری شمارش می‌شود) دلائلی که مطرح کرده‌اند : به دلیل وجود یک دوربین کیفیت پائین در بالای ماز هشت‌بازوئی در زمان‌واقعی شناسائی این حالت‌ها نباید برروی آموزش خودکار تاثیر بگذارد. ولی باید حالت بی‌حرکتی و سکون در رت‌ربات شناسائی شود زیرا اگر تحریک در این مدت اعمال نشود این حالت طولانی می‌شود. روش‌های پیشنهادی در این حالت شناسائی گوشه‌های تصویر و بررسی تغییرات آن در فریم‌های مختلف می‌باشد که توانسته‌اند با دقت ۹۷/۱٪ این حالت را شناسائی کنند [۱۸].

### ۴-۲-۲-۲- طرح وظائف یادگیری

براساس طرح وظائف یادگیری تصمیم‌گیری مرتبط ارسال می‌شود. محیط آموزش یک ماز هشت‌بازوئی است (شکل۱۳). بنابراین می‌توان وظائف آموزشی را به هشت وظیفه تقسیم کرد که حرکت بین بازوها می‌باشد.

### ۴-۲-۲-۳- ارزیابی وضعیت یادگیری

---

[1] Grooming

طرح وظائف یادگیری یک راهنما برای رت‌ربات در زمانی که رت حرکات نرمال دارد می‌باشد. هنگامی که رت رفتارهای غیرطبیعی از خود نشان می‌دهد سیستم ارزیابی وضعیت یادگیری شروع به پردازش می‌کند و براساس سرعت حرکت فعلی رت‌ربات یا مدت زمان بی‌حرکتی تصمیم می‌گیرد.

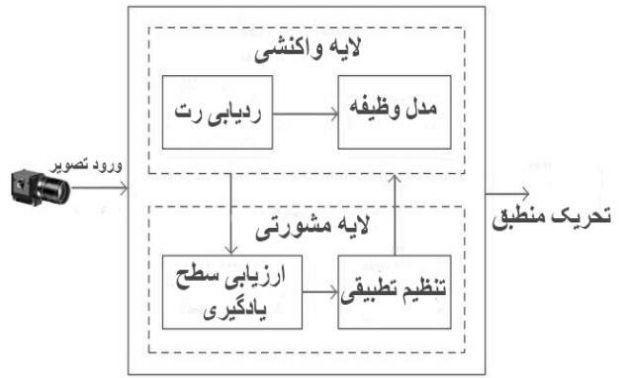

**شکل(۱۲)- چارچوب کلی آموزش خودکار**

اگر دیده شود که بی‌حرکتی در یک آزمایش بیش از حد تکرار شده است به این معنی است که تحریک پاداش کم اثر است و باید افزایش یابد. علاوه براین، اگر حرکت رت‌ربات بیش از حد سریع باشد به این معنی است که تحریک بیش از حد زیاد است و ممکن است رت‌ربات به درستی هدایت نشود و دستورهای آموزشی ( روبه‌جلو ، چرخش به چپ و راست) را اشتباه انجام دهد و در این حالت باید شدت پاداش کاهش یابد. در این مدل، تعداد رفتار سکون و سرعت حرکت رت‌ربات به طور پیوسته ثبت می‌شود[۱۸].

### ۴-۴-۲-۲- تنظیم تطبیقی پاداش

تنظیم تطبیقی به طور پیوسته اتفاق نمی‌افتد ولی امری مهم در آموزش اتوماتیک می‌باشد. این سیستم تنظیم کننده مدل ارزیابی وضعیت یادگیری را براساس دو قانون نوشته شده توسط متخصص آموزش ارائه می‌دهد. دو قاعده به شرح زیر است : اولاً، تعداد رفتارهای بی‌حرکتی و سکون که افزایش سطح تحریک پاداش را به‌همراه دارد. دوماً، درصورت بالا بودن سرعت حرکت، سطح پاداش باید کاهش یابد. براساس پاسخ تحریک هرکدام از رت‌ربات ها تحریک پاداش را به سطوح مختلف تقسیم می‌کنند [۶۱]. هر کدام از رت‌ربات ها یک محدوده مورد قبول تحریک پاداش دارند و در این رنج مورد قبول، فرض بر آن است که سطح بالاتر هیجان بیشتری برای رت‌ربات خواهد داشت یعنی با افزایش سطح تحریک در این رنج پاسخگوئی بهتری را شاهد بوده اند. از آنجائی که توانایی انسان برای هدایت رت‌ربات به

اثبات رسیده‌است، یک روش مرسوم برای تحقق بخشیدن به هدایت اتوماتیک رت‌ربات با مدلی ریاضی با یادگیری و تقلید از هدایت انسانی رت‌ربات پیشنهاد شده است. این روش به‌دور از روش‌های مرسوم که سعی در بیان صریح منطق کنترل دارند تعریف شده‌است. مدل ناوبری براساس یادگیری از هدایت انسان بنا شده و اولین روش برای هدایت اتوماتیک بیوربات می‌باشد.

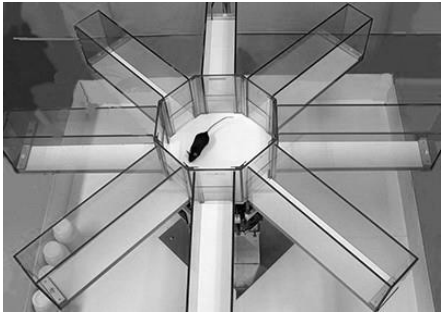

**شکل(۱۳)- ماز هشت بازوئی**

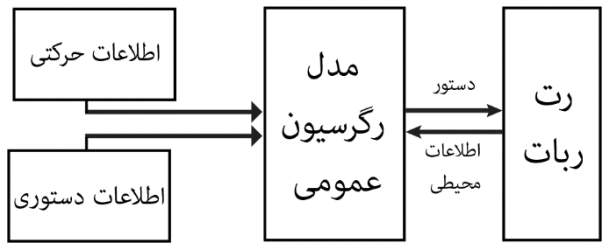

**شکل(۱۴)- نحوه شکل‌گیری مدل رگرسیون عمومی**

دراین روش از مدل ریاضی شبکه‌های عصبی رگرسیون عمومی[1] به عنوان مدل ریاضی استفاده شده‌است. در مرحله اول، رت‌ربات به‌وسیله اپراتور انسانی هدایت می‌شود و دستورات کنترل و تصاویر نظارتی ضبط و حرکت رت‌ربات از تصاویر استخراج شده و به‌عنوان ورودی به‌مدل اعمال می‌شود.

علاوه بر حرکت‌ها، داده های دیگری نیز به‌عنوان ورودی گرفته شده‌اند. شرح داده های ورودی در جدول ۲ آمده است. دستورات متناظر هماهنگ و براساس برچسب ها کدگذاری می‌شوند. با این موارد، مدل شروع به پیکربندی شبکه می‌کند. مرحله دوم، مدل رگرسیون عمومی برای پیش‌بینی دستورات به‌عنوان فرمان کنترل، داده‌های ورودی را طبقه‌بندی می‌کند و در نتیجه هدایت رت‌ربات بدون اپراتور انسانی انجام می‌پذیرد. نمودار الگوریتم در شکل ۱۴ نشان داده شده است. دوربین برای نظارت برکل صحنه آزمایش مهیا شده و تصاویر ویدیویی به‌صورت زمان‌واقعی به‌سیستم نرم‌افزاری منتقل می‌شود. حرکت رت‌ربات توسط ماژول تجزیه و تحلیل حرکت با تکنیک‌های بینائی ماشین ردیابی و استخراج می‌شود و باتوجه به پارامترها، ماژول تصمیم‌گیری، دستورات را برای رت‌ربات تولید می‌کند.

---

[1] General Regression Neural Network

این ماژول توسط الگوریتم مدل سازی رگرسیون عمومی کار می‌کند. دستورات محاسبه شده و منتقل شده به کوله‌پشتی، توسط تحریک‌الکتریکی به مغز اعمال می‌شود[۱۵].

### ۵-۲-۲-۴- مدارهای تحریک الکتریکی

سیستم رت‌ربات شامل دو مدول سخت افزاری تحریک کننده و سیستم بی‌سیم است. سیستم بی‌سیم از دو بخش ثابت و بخش سیار تشکیل شده است. بخش ثابت شامل یک لپ‌تاپ، یک ریزپردازنده، فرستنده وگیرنده است و بخش سیار از یک سر و کوله‌پشتی تشکیل شده است. مدار تحریک کننده، پالس-های تحریک را تولید می‌کند. پردازنده‌ی اصلی در تحریک کننده یک ریزپردازنده است که سرعت بالا، اندازه‌ی کوچک و مصرف توان کمی دارد. این خصوصیات این پردازنده را برای استفاده در کوله‌پشتی روی رت، مناسب می‌کند. این پردازنده

**جدول(۲)- پارامترهای ورودی به مدل**

| واحد | پارامتر |
|---|---|
| پیکسل | $\Delta X$ |
| پیکسل | $\Delta Y$ |
| درجه | جهت |
| پیکسل | فاصله (D) |
| درجه | زاویه چرخش(TA) |
| درجه | زاویه انحراف(OA) |
| میلی ثانیه | زمان گذشته از آخرین زمان |

دو مبدل آنالوگ به دیجیتال ۱۲ بیتی دارد که خروجی هایی برای تولید شکل موج منظم ایجاد می کند. پالس های تحریک الکتریکی از دو مبدل آنالوگ به دیجیتال پردازنده‌ی بالا صادر می شود تا برای کنترل یک مدار درایور ولتاژ ثابت و یک مدار درایور جریان ثابت و در نتیجه تولید پالس استفاده شود. با استفاده از سه مدار سوئیچ آنالوگ، تحریک کننده به عنوان تولید کننده‌ی پالس که خروجی آن پالس های جریانی یا ولتاژی هستند، عمل می کند. دامنه های پالس های خروجی،

متغیر هستند بنابراین، تحریک کننده می تواند سیگنال هایی با شکل موج های مختلف برای رفع نیازهای متفاوت، تولید کند.

## ۵- رویکردهای جدید

مطالعات در سال‌های اخیر بر روی دو محور اصلی بنا شده‌است. اولین محور بر مبنای طراحی سیستم هدایت بدون نیاز به آموزش اولیه می‌باشد دومین محور نوآوری بر مبنای روش‌ها و اصول مدولاسیون عصبی است (تحریک نوری و اپتوژنتیک) که تا به حال نتایج آن رضایت‌بخش نبوده است. اکثر پژوهش‌های صورت پذیرفته در سال‌های اخیر بر محوریت معرفی نواحی کارآمد در امر هدایت می‌باشد که بتوان به وسیله این نواحی فاز آموزش اولیه را کوتاه نمایند و همچنین به کارایی بهتری در هدایت اتوماتیک برسند. مطالعات سال اخیر میلادی بر روی تحریک مسیر نیگرواستریاتال [۳۹] و هسته تالاموس-پسترامدیال‌شکمی[۷۸] افق جدیدی در امر هدایت و جهت‌دهی رت متصور شده اند. محققان امیدوارند در آینده بدون هیچ آموزش اولیه ای سیستم هدایت کارآیی براساس تحریک این نواحی طراحی نمایند و سیستم های هوشمندی از ترکیب هوش مصنوعی و هوش بیولوژیکی پدید آورند که توانایی بهتری نسبت به هوش مصنوعی و بیولوژیک به‌تنهایی داشته‌باشد و راه‌اندازی این سیستم در کوتاه‌ترین زمان امکان‌پذیر باشد. در ادامه شرح مختصری از رویکردهای جدید را آورده‌ایم. مسیر نیگرواستریاتال به عنوان جدیدترین ناحیه‌ای معرفی شده‌است که تحریک این ناحیه جهت‌دهی و پاداش توام می‌باشد و نشان داده‌اند که تحریک الکتریکی نیگرواستریاتال در موش نیاز به آموزش اولیه همانند MFB ندارد [ ۳۹]. یک روش کنترل جهت‌یابی جدید از طریق تحریک ناحیه شکمی هسته تالاموس ارایه شده است که از طریق تحریک مصنوعی هسته تالاموس پسترامدیال-شکمی می‌باشد. تحریک هسته تالاموس پسترامدیکال شکمی باعث می شود که عنصر حرکات ارادی از مکانیزم حرکت حذف شود. در مقایسه با روش قبلی (تحریک قشر بارل) این روش پروسه آموزش را حذف کرده است. یادگیری فضایی در سیستم-های مبتنی بر کامپیوتر می‌تواند بر آموزش و تصمیم‌گیری یک عنصر بیولوژیکی ( برای مثال رت) تاثیر بگذارد [۷۸]. وقتی که هوش بیولوژیکی و مصنوعی به‌هم وصل می‌شود یک ماهیت جدیدی از هوش را پدید می‌آورد که هوش سایبورگ نام‌گذاری کرده اند. یک سری وظایف آموزشی برای بررسی توانایی

---
[1] Ventral Posteromedial Thalamic Nucleus

یادگیری رت‌ربات در یک ماز پیچیده طراحی کرده‌اند. سه نوع قاعده هدایت در مقالات مطرح شده است:
۱- قواعد یادگیری [۲۰].
۲- قواعد نرم‌افزاری [۶۲].
۳- قواعد ترکیبی و انتقال [۶۳].

**الف) قواعد یادگیری**

کامپیوتر مسیر هدایت رت را دنبال می‌کند و واحدهای سپری‌شده را ذخیره می‌نماید. با استفاده از الگوریتم‌های تقویت‌شده (برای مثال الگوریتم یادگیری Q)، کامپیوتر ماز را توسط برنامه‌ای پویا بعد از اولین آزمایش حل می‌کند و براساس جداول Q- نقشه پاداش را طی می‌نماید. کامپیوتر افزایش تحریک پاداش را از نقشه پاداش استخراج می‌کند [۲۰, ۶۳].

**ب) قواعد نرم‌افزاری**

دوربین کوچکی بر سر رت‌ربات کار گذاشته‌اند و علامت‌ها را شناسائی می‌کند و کامپیوتر براساس علامت‌هایی که از محیط آزمایش، جهت را مشخص می‌کند و در نقطه تصمیم‌گیری اقدام به تحریک ناحیه MFB می‌کند و عمل هدایت به این صورت انجام می‌شود [۶۲].

**ج) قواعد ترکیبی و انتقال**

کامپیوتر ترکیبی از قواعد افزایش پاداش و پیروی از علامت را با هم در این مدل اعمال می‌نماید به صورتی که ابتدا دوربین کار گذاشته‌شده بر روی سر، علامت را تشخیص می‌دهد و کامپیوتر نیز سطح تحریک MFB را افزایش می‌دهد و رت را در امتداد جهت مشخص شده توسط علامت هدایت می‌کند. واسط‌های مغز-مغز۱، باعث پیوندِ مغز افراد به اطلاعات با پهنای‌باند گسترده بدون هیچ‌گونه زبان رسمی یا فعالیتی می‌شود [۶۴]. علاوه بر این واسط‌های مغز-مغز برای ارتباط با حیوانات به خصوص حیوانات اموزش ندیده مفید می‌باشد [۶۵]. اولین واسط مغز-مغز، سیگنال‌های مغزی را از موش رمزگذار۲ برای مطابقت با سیگنال‌های مغزی کورتیکال موش رمزگشا۳ ارسال کرد که اطلاعات رفتاری انها به اشتراک گذاشته شد [۶۵]. درسال‌های اخیر ارتباطات انسان-انسان با استفاده ازواسط های مغز-مغز غیرتهاجمی گزارش شده است [۶۶, ۶۷]. تمام سیستم‌های قبلی نیازمند بیهوشی و یا اموزش قبلی بودند که برای رفع این محدودیت یک سیستم جدید واسط مغز-مغز که انسان قادر به کنترل رفتارهای حیوان می‌باشد طراحی شد که

ویژگی این سیستم عدم نیاز به اموزش قبلی بود. این سیستم جدید از واسط مغز-کامپیوتر که ورودی را ازپتانسیل‌های برانگیخته بینایی۴ دریافت می‌کند و واسط کامپیوتر-مغز که مسیر نیگرواستریاتال موش را هدف قرار می‌دهد ساخته شده است. الکترود در مسیر نیگرواستریاتال رت کاشته می‌شود و پالس‌تحریک را اعمال می‌کنند. در صورتی که رت‌ربات چرخشی به اندازه سی‌درجه انجام می‌داده است آزمایش مورد تایید قرار می‌گرفت که در شکل ۱۵ کاملا مشهود است. در این آزمایش نیز زمان انجام تسک مورد مهمی است که میانگین سرعت محاسبه شده ۱/۹ متر در دقیقه بوده است [۳۹]. به غیر از تحریک الکتریکی قشر حسی پیکری، روش‌هایی برای ایجاد نشانه‌دهی هدایت در سایر اشکال تحریک نظیر تحریک‌نوری وجود دارد. هابر و همکاران با ریز تحریک‌نوری قشر بارل توانسته‌اند رفتاری را به موش در حرکت آزادانه آموزش دهند و موش آموزش‌دیده تحریک نوری را تشخیص دهد [۵۳]. در زمان‌هایی که تحریک‌های تهاجمی ضرورتی ندارد، روش‌های غیر تهاجمی برای نشانه به حیوانات با استفاده از تحریک اپتوژنتیک۵ اعمال می‌شوُد. سلچیدهاناندام و همکاران ناحیه بارل در رت را بوسیله روش اپتوژنتیک تحریک نموده‌اند و دریافته‌اند که آموزش اپتوژنتیک مبتنی بر پاداش می‌تواند در کنترل رفتار رت مورد استفاده قرار بگیرد [۵۴]. کنترل حرکتی اپتوژنتیک حیوانات بینشی برای شبکه‌های عصبی و نوروآناتومی کاربردی که شامل فعالیت ناحیه‌حسی حرکتی، پاداش، احساسات و یادگیری ترس می‌باشد فراهم کرده است [۵۳, ۴۷, ۶۸, ۶۹] بسیاری از نواحی مغز دارای پتانسیل برای کنترل حرکت حیوانات هستند با این حال تحقیقات کمتری با استفاده از اپتوژنتیک برای طراحی این مکانیزم انجام شده است. تیم‌های تحقیقاتی‌ای قشر حرکتی را بصورت نوری و مستقیم تحریک می‌نمودند [۷۰]. تیم ماتسوحیرو قسمت قدامی قشرحرکتی را تحریک کرده و انقباضات عضله را در اطراف چپ گردن و دست بالایی و چرخش به چپ ثبت کردند [۶۸]. کنترل حرکتی چرخش حتی با وجود نقشه‌های قشرحرکتی چالش برانگیز می‌باشد [۷۱]. تحریک ناحیه حسی حرکتی حیوان باعث بروز رفتار چرخش می‌گردد. حیوانات با ویسکر بطور مثال جوندگان از این حس برای جمع‌آوری اطلاعات برای جهت‌یابی استفاده می‌کنند [۷۲]. با نشانه‌های حسی و آموزش‌ها برپایه پاداش قادر به

---

کنترل و هدایت حیوانات می‌گردند [۱۵, ۷۳]. مطالعات برروی مخچه حرکات جنبشی خوبی را نشان می‌دهد و همبستگی بین مدارهای حسی حرکتی نشان می‌دهد که پتانسیل کنترل حرکات حیوانات را دارد [۷۴].

## ۶- جمع‌بندی

در این مقاله مروری بر مطالعات مرتبط با سیستم‌های جهت‌دهی و هدایت رت توسط تحریک‌الکتریکی صورت پذیرفته‌است. روش‌های متعددی برای مدالیته فعالیت‌های عصبی مانند روش الکتریکی، اپتیکی، مکانیکی، حرارتی و اپتوژنتیک وجود دارد. در میان این‌ها، مدالیته به‌وسیله تحریک الکتریکی روش اصلی برای برنامه‌های تحقیقاتی و بالینی است و عمدتاً به این دلیل است که سیستم تحریک‌الکتریکی عصبی را می‌توان به‌راحتی پیاده‌سازی و کوچک‌سازی کرد. با این وجود، به‌منظور تحریک‌الکتریکی به‌طور مؤثر به بافت‌ها یا سلول های عصبی، عوامل و پارامترهای متعددی باید در نظر گرفته‌شود. در ابتدا نواحی تحریک الکتریکی را در سه رویکرد متفاوت طبقه‌بندی کرده‌ایم این سه رویکرد پاداش، حسی‌پیکری و موتوری می‌باشد. روش‌های مبتنی بر تحریک پاداش و حسی‌پیکری روش‌های مرسوم

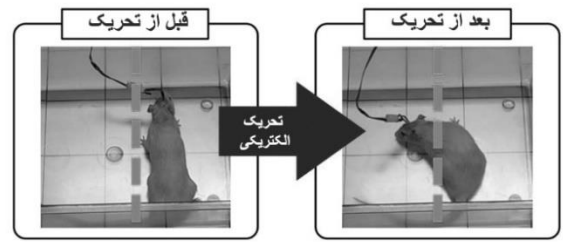

**شکل(۱۵)- تغییر در زاویه چرخش رت‌ربات در پاسخ به تحریک الکتریکی نیم کره راست مغز مسیر نیگرواستریاتال[۳۹].**

در امر هدایت است. سپس اصول و روش‌های تحریک الکتریکی مورد بررسی قرار گرفت. روش‌های هدایت و آموزش رت‌ربات به‌وسیله تحریک الکتریکی دو رویکرد متفاوت دارد در رویکرد اول، آموزش و هدایت رت‌ربات به‌صورت دستی صورت می‌پذیرد و در رویکرد دوم، با الگوبرداری از روش قبل، آموزش و هدایت اتوماتیک را معرفی کرده اند و سعی نموده‌اند به وسیله روش‌های ریاضی هدایت اتوماتیک را شکل دهند. در آخر به بررسی رویکردهای جدید در امر هدایت پرداخته‌ایم که در مراحل اولیه بررسی و تکامل می‌باشند.

## ۷- سپاس‌گزاری